\documentclass[structabstract]{aa}  
\usepackage{graphicx}
\usepackage{multirow}
\usepackage{natbib}
\usepackage{eqnarray}
\usepackage{txfonts}
\begin{document}

\title{Evolutionary aspects and north-south asymmetry of soft X-ray flare index during solar cycles 21, 22, and 23}

   \author{Bhuwan Joshi
          \inst{1,3}
          \and
          R. Bhattacharyya
          \inst{1}
          \and
          K. K. Pandey
          \inst{2}
          \and
          Upendra Kushwaha
          \inst{1}
          \and
          Yong-Jae Moon
          \inst{3}
          }

   \institute{Udaipur Solar Observatory, Physical Research Laboratory, Udaipur 313 004, India\\
              \email{bhuwan@prl.res.in, ramit@prl.res.in, upendra@prl.res.in}
         \and
             Department of Astronomy, Osmania University, Hyderabad 500 007\\
             \email{pandeyou@yahoo.co.in}
         \and
            School of Space Research, Kyung Hee University, Yongin, Gyeonggi-Do, 446-701, Korea
            \email{moonyj@khu.ac.kr}   
            }

   \date{Received ; accepted }

   \abstract
  {}
   {In this paper, we investigate the temporal evolution and north-south (N-S) asymmetry in the occurrence of solar flares during cycle 21, 22, and 23, and compare the results with traditional solar activity indices.}
   {The flare activity is characterized by a soft X-ray (SXR) flare index,  which incorporates information about flare occurrences during a selected interval along with the peak intensity of individual events.}
   {The SXR flare index correlates well with other conventional parameters of solar activity. Further, it exhibits a significantly higher correlation with sunspot area over sunspot number, which suggests the variations in sunspot area to be more closely linked with the transient energy release in the  solar corona. The cumulative plots of the flare index indicate a slight excess of activity in the northern hemisphere during cycle 21, while a southern excess clearly prevails for cycles 22 and 23. The study reveals a significant N-S asymmetry, which exhibits variations with the phases of solar cycle. The reliability and persistency of this asymmetry significantly increases when the data is averaged over longer periods, while an optimal level is achieved when data is binned for 13~Carrington rotations. The time evolution of the flare index further confirms 
evolution of dual peaks in solar cycles during the solar maxima and violation of Gnevyshev-Ohl rule for the pair of solar cycles 22 and 23.}
{The SXR flare index in the northern and the southern hemispheres of the Sun exhibits significant asymmetry during the evolutionary phases of the solar cycle, which implies that N-S asymmetry of solar flares is manifested in terms of the flare counts as well as the intensity of flare events.}

\titlerunning{Evolutionary aspects and north-south asymmetry in solar activity}
\maketitle

%

\section{Introduction}
The Sun displays a multitude of magnetic activities, collectively referred as the solar activity, across a wide range of spatial, temporal, and energy scales. These activities while being time dependent are also spatially localized on the photosphere and in the solar atmosphere. The quasi-periodic variation of $\sim$11 years is known to be the most fundamental manifestation of solar activity characterizing a systematic pattern in solar magnetism over long timescales. In short-term variations, the 27-day periodicity is the most prominent and is attributed to the solar rotation. These periodic
variations are reflected in many solar activity indices related to photospheric, chromospheric, and coronal activity.

The solar active phenomena in their different manifestations (e.g., sunspots, flares, prominences, etc.) are known to have nonuniform spatial distributions at the corresponding region of occurrences. Further, it has also been recognized that there is a statistical imbalance in these occurrences of solar activity between the northern and the southern hemispheres of the Sun
when averaged over a suitable timescale. This phenomenon, intrinsic to the active Sun, is known as the N-S asymmetry, the prediction of which may pose a challenge for a successful solar dynamo model. However, the N-S asymmetry exhibits temporal evolution within a given solar cycle along with significant variations in different solar cycles. Investigations of the distribution of these activity features  over the solar disk and their 
evolution with solar cycles are then of fundamental importance, 
leading to a better understanding of the origin and evolution of the solar magnetism. 

The existence of N-S asymmetry has been investigated utilizing several activity indices like sunspot number, sunspot area, flare, filament, magnetic flux, coronal mass ejection, etc. \citep[e.g., ][]{howard1974,garcia1990,verma1993,temmer2001,joshi2005aa,joshi2006, zharkov2005, temmer2006,li2009,joshi2009,gao2009,zharkov2011}. These studies reveal the real nature of asymmetry in different manifestations of the solar magnetic fields. \cite{howard1974} investigated the N-S distribution of solar magnetic flux for the period 1967--1973 and found that about 95\% of the total magnetic flux of the Sun is confined to latitudes below 40$^{\circ}$ in both hemispheres. It was also found that the total magnetic flux in the north exceeds that in the south by 7\%. \cite{roy1977} studied the N-S distribution of major flares (during the period 1955--1974), sunspot magnetic configuration (during the period 1962--1974), and sunspot area (during the period 1955--1974) and found that the northern hemisphere dominates the southern in all these categories. \cite{garcia1990} performed a systematic study of N-S distribution of soft X-ray flares (class $\geq$ M1) during solar cycles 20 and 21. This study clearly reveals that the spatial distribution of flares varies within a solar cycle such that the preponderance of flares occurred in the north during the early
part of the cycle and then moves toward south as the cycle progresses.
\cite{Li1998} investigated the N-S distribution
of soft X-ray flares during solar cycle 22 and found the dominance of southern hemisphere. \cite{verma2000} examined the latitudinal distribution of solar active prominences (SAP) for the period 1957--1998 and detected a significant N-S asymmetry that does not seem to have relation with
the maximum or minimum phases of solar cycle. \cite{temmer2002} found significant N-S asymmetry in sunspot numbers
during the period 1975--2002. They analyzed the hemispherical asymmetry in context of rotational behaviors in the northern and southern hemispheres and concluded that the magnetic field systems originating in the two hemispheres to be only weakly coupled. 

A meaningful statistical analysis of any observation requires a data set optimal in terms of timescales and physical processes inherent to that observation. Based on this, in this paper we present a statistical analysis of solar X-ray flares observed by Geostationary Operational Environment Satellites (GOES). The novelty of this analysis lies in the characterization of flare activity by soft X-ray flare index instead of the traditional flare counts. The flare activity has been compared with several other activity parameters that occur in different atmospheric layers of the Sun. We have further examined temporal evolution of flare activities along with their N-S asymmetry and discuss the implications of the results toward a better understanding of the solar cycle. 

The paper is organized as follows. In section~\ref{sec_FI}, we present a plausible rationale for choosing the soft X-ray flare index as 
a favored solar activity indicator over other conventional counterparts. The data analysis and results are presented in section~\ref{sec_analysis}. In section~\ref{sec_dissus}, we discuss the results of our analysis in detail. 



\section{Soft X-ray flare index}
\label{sec_FI}

\begin{table*}
\caption{Occurrence of soft X-ray flares during solar cycle 21, 22, and 23.} 
\label{table:1}
\centering
\begin{tabular}{c |r r| r r |r r}
\hline\hline
Class&
      \multicolumn{2}{c}{Cycle 21} &
      \multicolumn{2}{c}{Cycle 22} &
      \multicolumn{2}{c}{Cycle 23} \\ \cline{2-7}
    &Counts & $FI_{SXR}$ &Counts & $FI_{SXR}$ &Counts & $FI_{SXR}$\\
\hline
X & 148 (1.4 \%)  & 33870 (33.8 \%)  & 144 (1.3 \%)  & 31130 (30.8 \%) & 107 (0.9 \%)   & 27820 (35.3 \%)  \\
M & 1681 (15.6 \%)& 39265 (39.1 \%) & 1673 (14.6 \%)& 40591 (40.2 \%)& 1060 (8.8 \%)  & 25799 (32.7 \%)\\
C & 8033 (74.6 \%)& 26670 (26.6 \%) & 7821 (68.0 \%)& 28177 (28 \%) & 7259 (59.9 \%) & 23395 (29.7 \%) \\
B & 900 (8.4 \%)  & 464 (0.5 \%)  & 1857 (16.1 \%)  & 1036 (1 \%)  & 3674 (30.4 \%)& 1776 (2.3 \%)\\
\hline
\end{tabular}
\tablefoot{
The events of soft X-ray flares over the solar cycle are characterized by their counts as well as soft X-ray flare index ($FI_{SXR}$). The value in brackets denotes the contribution of the corresponding parameter out of its total value over that particular solar cycle. The $FI_{SXR}$ has been rounded up to the nearest integer.
}
\end{table*}

\begin{table*}
\caption{Start and end times of solar cycles 21, 22, and 23 in terms of Carrington rotation (CR) and calendar units.}             
\label{table:2}      
\centering                          
\begin{tabular}{ c l l}        
\hline\hline                 
Solar cycles & Start & End \\    
\hline                        
Cycle 21 & CR 1642 (June 1976) & CR 1779 (August 1986) \\     
Cycle 22 & CR 1780 (September 1986) & CR 1909 (April 1996) \\
Cycle 23 & CR 1910 (May 1996) & CR 2080 (December 2008) \\
\hline                                   
\end{tabular}
\end{table*}

\begin{figure*}
\sidecaption
\includegraphics[width=12cm]{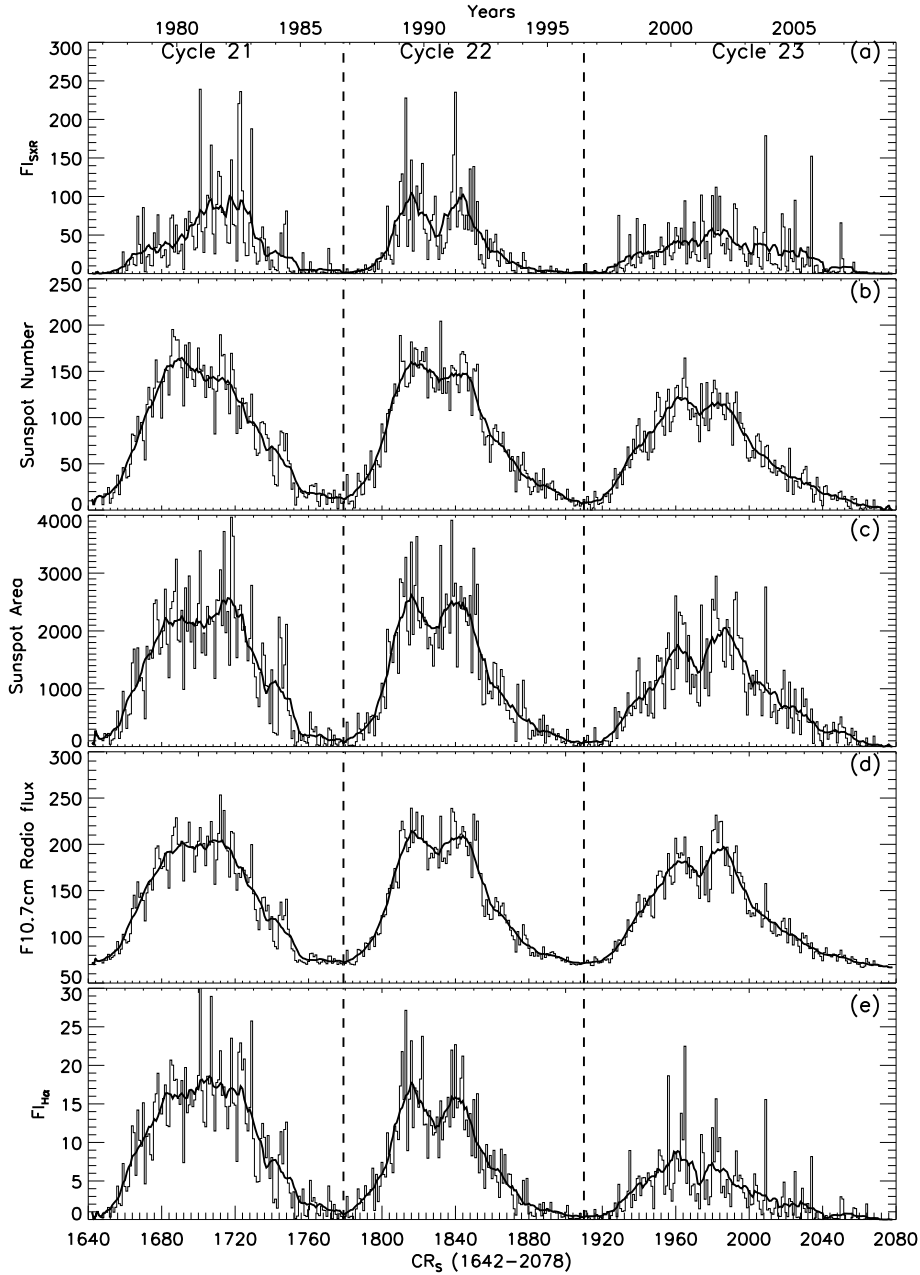}
\caption{Temporal evolution of $FI_{SXR}$, sunspot number, sunspot area, F10.7, and  $FI_{H\alpha}$  (from top to bottom panels) during solar cycles 21, 22, and 23. Different plots represent values of corresponding solar activity parameter averaged over a Carrington rotation (CR). The X-axis is labeled in the units of Carrington rotation as well as calendar year. The vertical dashed lines indicate the minimum phase of solar cycles. The smoothed curve in each plot indicates the 13-point running averages.
}
\label{Fig1}
\end{figure*}

\begin{figure*}
\sidecaption
\includegraphics[width=12cm]{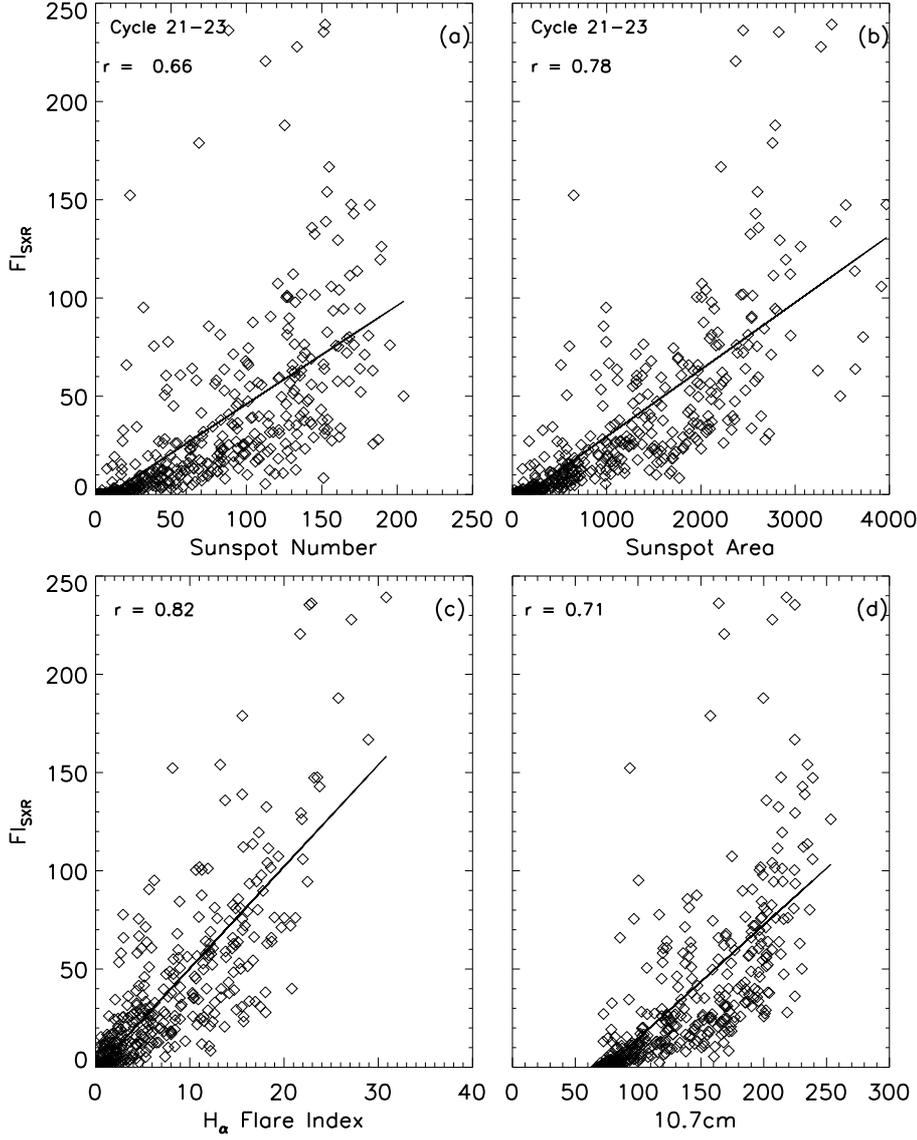}
\caption{Scatter plots showing correlation of sunspot number, sunspot area, $FI_{H\alpha}$, and F10.7 with $FI_{SXR}$. The value of correlation coefficient ($r$) is mentioned in each panel.}
\label{Fig2}
\end{figure*}                     

\begin{figure*}
\sidecaption
\includegraphics[width=12cm]{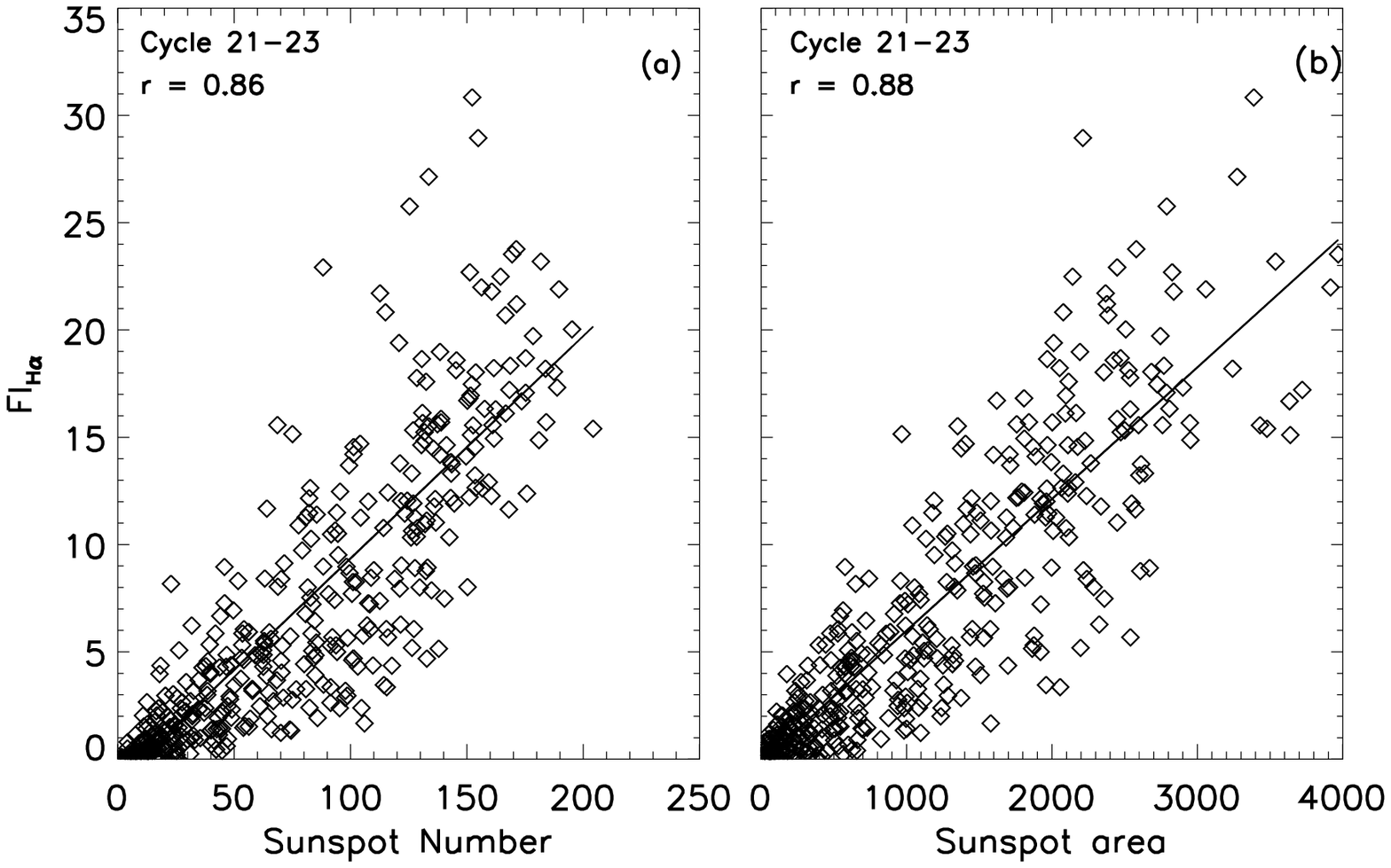}
\caption{Scatter plots showing correlation of $FI_{H\alpha}$ with sunspot number and sunspot area. The value of correlation coefficient ($r$) is mentioned in each panel.}
\label{Fig3}
\end{figure*}

We consider the observations of soft X-ray flares  taken by Earth orbiting GOES satellite. In the GOES classification scheme, the category of a flare is denoted by a letter (A, B, C, M, or X) corresponding to the peak soft X-ray flux (ranging from~$<$10$^{-7}$ $Wm^{-2}$~to~$>$10$^{-4}$ $Wm^{-2}$), observed by 1$-$8~{\AA} channel along with a number (1$-$9) that acts as a multiplier. Therefore, the class of a flare directly provides information about the peak intensity of emission. Moreover, this X-ray classification of solar flares being based on the denary logarithmic scale, allowed \cite{antalova1996} to define the soft X-ray flare index ($FI_{SXR}$) by weighing the SXR flares of classes C, M, and X as 1, 10, and 100, respectively (i.e., in units of $10^{-6}$ $Wm^{-2}$). In this manner, for example, the SXR flare index for individual events of class C7.3, M2.9, and X4.8 would be 7.3, 29.0, and 480, respectively.  In our analysis, we  also included the contribution of flares of class B in the calculation of daily $FI_{SXR}$. In view of flares less intense than C1 being generally detected at the minimum phase of a cycle when the background is low, such inclusions of B class flares render $FI_{SXR}$ a very sensitive and useful indicator for analyzing short-term and small-scale fluctuations in flare activity \citep[e.g.,][]{joshi2005soph,abramenko2005}, even during phases of lower
solar activity. In contrast, during the maximum phase $FI_{SXR}$ is more biased toward major flares and any weighted contribution from B-class flares, although present, is negligible.

The daily $FI_{SXR}$ is given by

\begin{equation}
\begin{array}{l}
FI_{SXR} = 0.1 \times \sum_{j=1}^{j=N_{B}} m_{j}^{B} + 1.0 \times \sum_{j=1}^{j=N_{C}} m_{j}^{C} \\
~~~~~~~~~~~~~+ 10.0 \times \sum_{j=1}^{j=N_{M}} m_{j}^{M} + 100.0 \times \sum_{j=1}^{j=N_{X}} m_{j}^{X},
\end{array}
\end{equation}

with $m^{B}$, $m^{C}$, $m^{M}$, $m^{X}$ as the digit multipliers and  $N_{B}, N_{C}, N_{M}$, $N_{X}$ are the daily counts for flares of class B, C, M, and X, respectively \citep{joshi2004}. The $FI_{SXR}$ computed using the above expression  
agreeably represents daily variations of SXR flare activity and can be a versatile parameter to examine the flare productivity over different temporal and spatial scales \citep[see, e.g., ][]{landi1998}. In contemporary studies, flare counts are used as an index of solar activity where equal weight is given to flares of different intensities. Since less energetic flares (e.g., flare of SXR class below M1 or H$\alpha$ Subflares) constitute a large fraction of total flares \citep[up to $\sim$85-90\%; see, e.g.,][]{joshi2005aa}, a straight forward use of flare counts  yields results that are susceptible to a bias toward less energetic flares. 
The use of $FI_{SXR}$ circumvent this bias as proper weight is given to flares of different classes depending on their peak intensities, effectively
providing an intensity weighted flare statistics associated with all the active regions of different sizes and complexities present over the solar disk during an 
observation period. In Table~\ref{table:1} we present counts of soft X-ray flares of different classes and corresponding  $FI_{SXR}$ over solar cycles 21, 22, and 23. The year wise list of SXR flares observed by GOES are published by the National Geophysical Data Center (NGDC\footnote{http://www.ngdc.noaa.gov/ngdc.html}).
Since the ensuing analysis also correlates $FI_{SXR}$ with other activity indicators, in the following we summarize them for easy reference. 

\begin{itemize}
\item {\bf Sunspot number:} Historically, the relative sunspot number $R$ is the most fundamental index of solar activity. Each isolated cluster of sunspots is termed a sunspot group, and it may consist of one or a large number of distinct spots. The relative sunspot number is defined as $R = K (10g + s)$, where $g$ is the number of sunspot groups and $s$ is the total number of distinct spots. The scale factor $K$ is an observer related correction factor, which depends upon the actual seeing conditions and the instrument used. The sunspot number separately for the northern and southern hemisphere of the Sun is made available by the National Geophysical Data Center (NGDC\footnote{http://www.ngdc.noaa.gov/ngdc.html}).

\item {\bf Sunspot area:} The Royal Greenwich Observatory (RGO) started compiling sunspot observations from May 1874 from a small network of observatories to produce a data set of daily sunspot area. The observatory concluded this data set in 1976 after the US Air Force (USAF) started compiling data from its own Solar Optical Observing Network (SOON). This work was continued with the help of the US National Oceanic and Atmospheric Administration (NOAA). The daily data of sunspot areas, expressed in units of millionths of solar hemisphere, is available at the NASA's Marshall Space Flight Center\footnote{http://solarscience.msfc.nasa.gov/greenwch.shtml}. 

\item {\bf H$\alpha$ flare index:} Solar flares are traditionally observed in H$\alpha$ filter by ground based observatories. To quantify daily flare activity on the basis of H$\alpha$ measurements, \cite{kleczek1952} introduced flare index as $FI_{H\alpha} = i \times t$,  where $i$ represents the intensity scale of importance and $t$ the duration (in minutes) of the flare \citep{atac1987,atac1998}. The flare index
  $FI_{H\alpha}$ is assumed to represent roughly the total energy emitted by the flares. The daily flare index is computed by Kandilli Observatory\footnote{http://www.koeri.boun.edu.tr/eng/topeng.htm}, Istanbul, Turkey using the final list of $H\alpha$ flares compiled by the NGDC.   

\item {\bf F10.7 solar radio flux:} The Sun emits radio energy with a slowly varying intensity. The radio flux at 10.7 cm (F10.7), which originates from atmospheric layers high in the Sun's chromosphere and low in its corona, changes gradually from day-to-day, in response to the number of sunspot groups on the disk. At 10.7 cm (i.e., 2800 MHz), radio intensity levels consist of emission from three sources: from the undisturbed solar surface, from developing active regions, and from short-lived enhancements above the daily level. Solar flux density at 10.7 cm has been recorded routinely by radio telescopes near Ottawa (February 14, 1947$-$May 31, 1991) and Penticton, British Columbia, since June 1, 1991 and is available through NGDC.

\end{itemize}

\section{Analysis and results}
\label{sec_analysis}
\subsection{Solar cycle evolution}

\begin{table*}
\caption{Correlation coffefficient ($r$) of soft X-ray flare index ($FI_{SXR}$) with sunspot number (SN), sunspot area (SA), 10.7 cm solar radio flux (F10.7), and H$\alpha$ flare index ($FI_{H\alpha}$).}  
\label{table:3}      
\centering                          
\begin{tabular}{ c c c c c c}        
\hline\hline                 
Cycle no. & N& $r$ ($SN/FI_{SXR}$) & $r$~($SA/FI_{SXR}$) & $r$ ($F10.7/FI_{SXR}$) &  $r$ ($FI_{H\alpha}/FI_{SXR}$) \\    
\hline                        
Cycle 21 & 138 & 0.60 & 0.73 & 0.67 & 0.83 \\
Cycle 22 & 130 & 0.76 & 0.84 & 0.82 & 0.90 \\
Cycle 23 & 169 & 0.58 & 0.74 & 0.64 & 0.81 \\
\hline                                   
\end{tabular}
\tablefoot{
Correlation is studied between CR averaged values of the corresponding parameters.~$N$ denotes number of data points for each cycle. For each case the p-value is $ < $ 0.0001, which together with $r$, indicates a very high correlations between $FI_{SXR}$ and other solar activity parameters.
}
\end{table*}

The temporal evolution of $FI_{SXR}$ during solar cycles 21, 22, and 23 is shown in Fig.~\ref{Fig1} (panel~a). The plot shows the average of $FI_{SXR}$ over a Carrington rotation (CR). We believe this representation of solar activity in terms of CR (i.e., rotational average) to be an apt choice since in any dynamo theory the differential rotation is responsible 
for stretching out the poloidal field lines to generate the toroidal field.  The toroidal to poloidal field conversion occurs through the Babcock-Leighton mechanism, which along with the meridional circulation completes the solar dynamo. The consequent topological complexity of the magnetic field then produces the variety of active phenomena observed at the outer and atmospheric layers of the Sun. Moreover, by averaging over a CR, we suppress the large fluctuations generated over short intervals due to the violent activity from a particular active region (AR), as a typical AR would decay substantially over a single Carrington rotation. In Fig.~\ref{Fig1}, we  also show the rotational average of sunspot number, sunspot area, F10.7, and $FI_{H\alpha}$ (panels b-e). Each panel of Fig.~\ref{Fig1} also shows the 13-point running average of the corresponding parameter.  In Table~\ref{table:2}, we summarize onset and end times of solar cycles 21-23 in terms of CR as well as calendar units. 

In Fig.~\ref{Fig2}, we show scatter plots between $FI_{SXR}$ and other indices of solar activity, viz., sunspot number, sunspot area, $FI_{H\alpha}$, and F10.7  with data from all 
the three cycles being considered. We find that $FI_{SXR}$ has good positive ($r$ $>$0.6) correlation with other activity parameters. In Table~\ref{table:3}, we list correlation coefficients between above mentioned parameters calculated for individual cycles. Since both $FI_{SXR}$ and $FI_{H\alpha}$ represent flare emission at different levels of solar atmosphere, it is worthwhile to compare how these two flare parameters correlate with sunspot activity. For this comparison, we depict correlation of $FI_{H\alpha}$ with sunspot number and sunspot area in Fig.~\ref{Fig3}.

\subsection{North--south asymmetry}

\begin{figure*}
\sidecaption
\includegraphics[width=12cm]{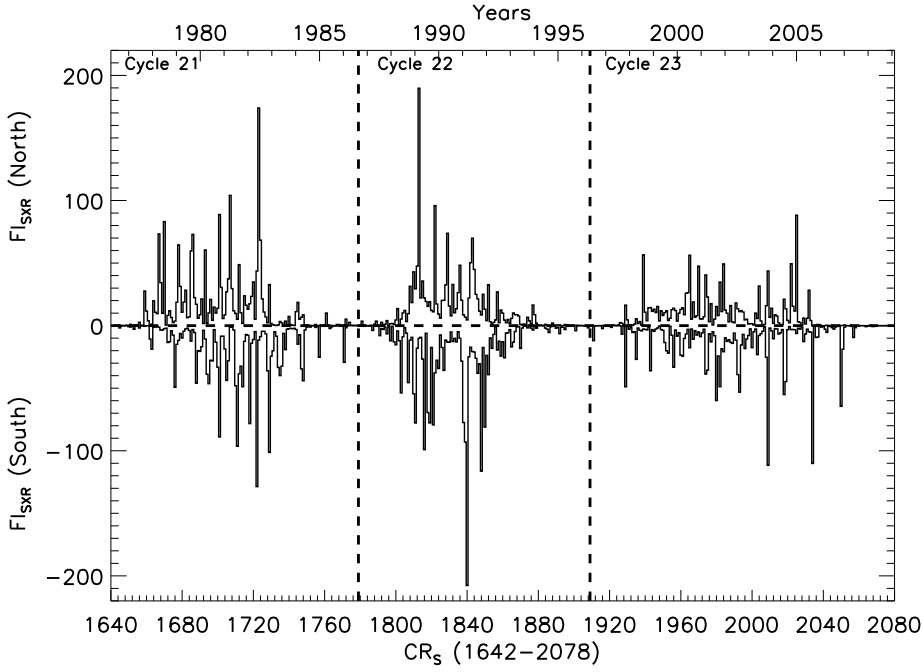}
\caption{Evolution of $FI_{SXR}$ in the northern and the southern hemispheres during solar cycle 21, 22, and 23. Note that X-axis is labeled in the units of Carrington rotation as well as calendar year. The vertical dashed lines indicate the minimum phase of solar cycles. }
\label{Fig4}
\end{figure*}

\begin{figure*}
\sidecaption
\includegraphics[width=12.5cm]{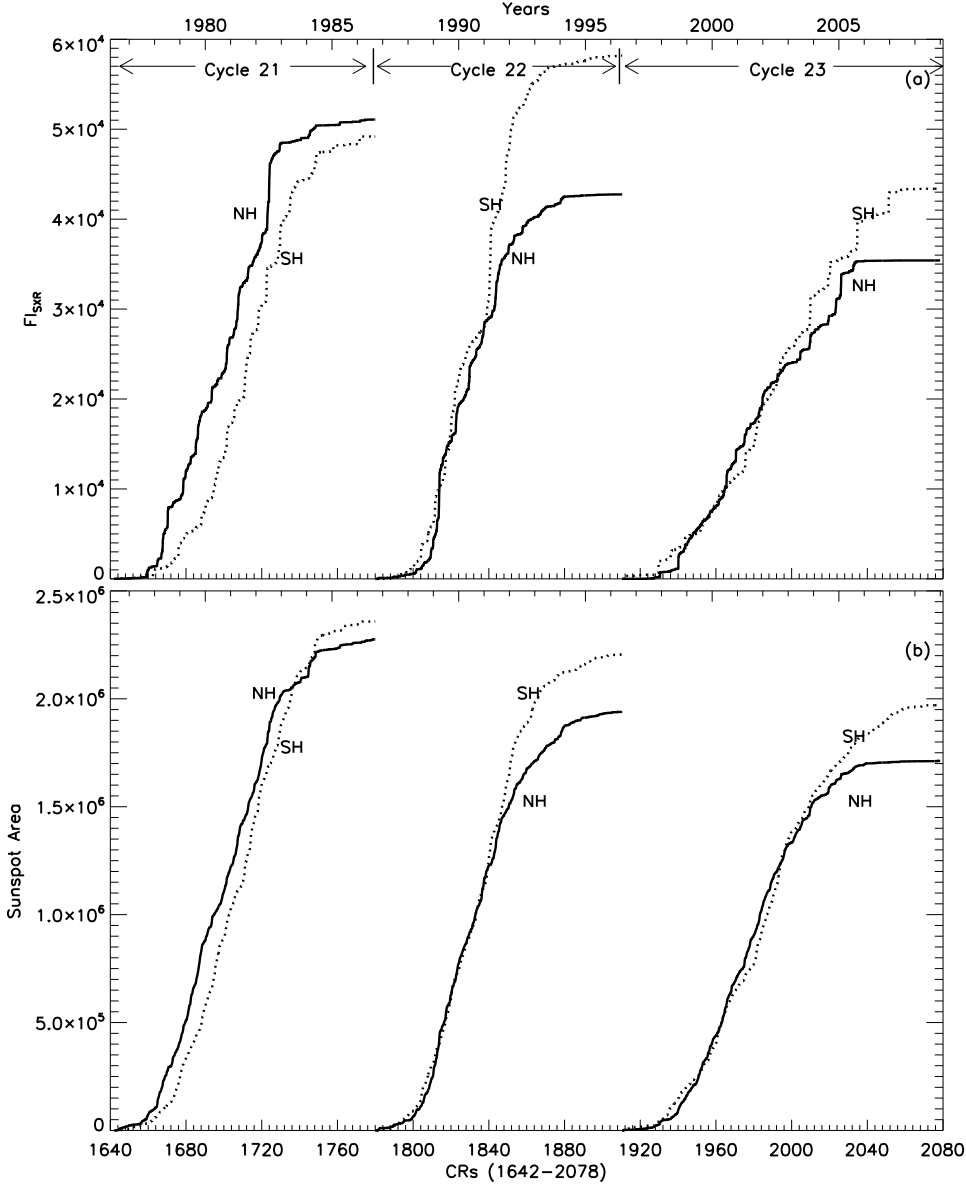}
\caption{Cumulative values of $FI_{SXR}$ (top panel) and sunspot area (bottom panel) for the northern (solid line) and the southern (dotted line) hemispheres during solar cycle 21, 22, and 23. In this representation, the vertical distance between the two curves (i.e., solid and dotted lines) at any time characterizes the north/south excess up to that epoch.}
\label{Fig5}
\end{figure*}

\begin{figure*}
\sidecaption
\includegraphics[width=12cm]{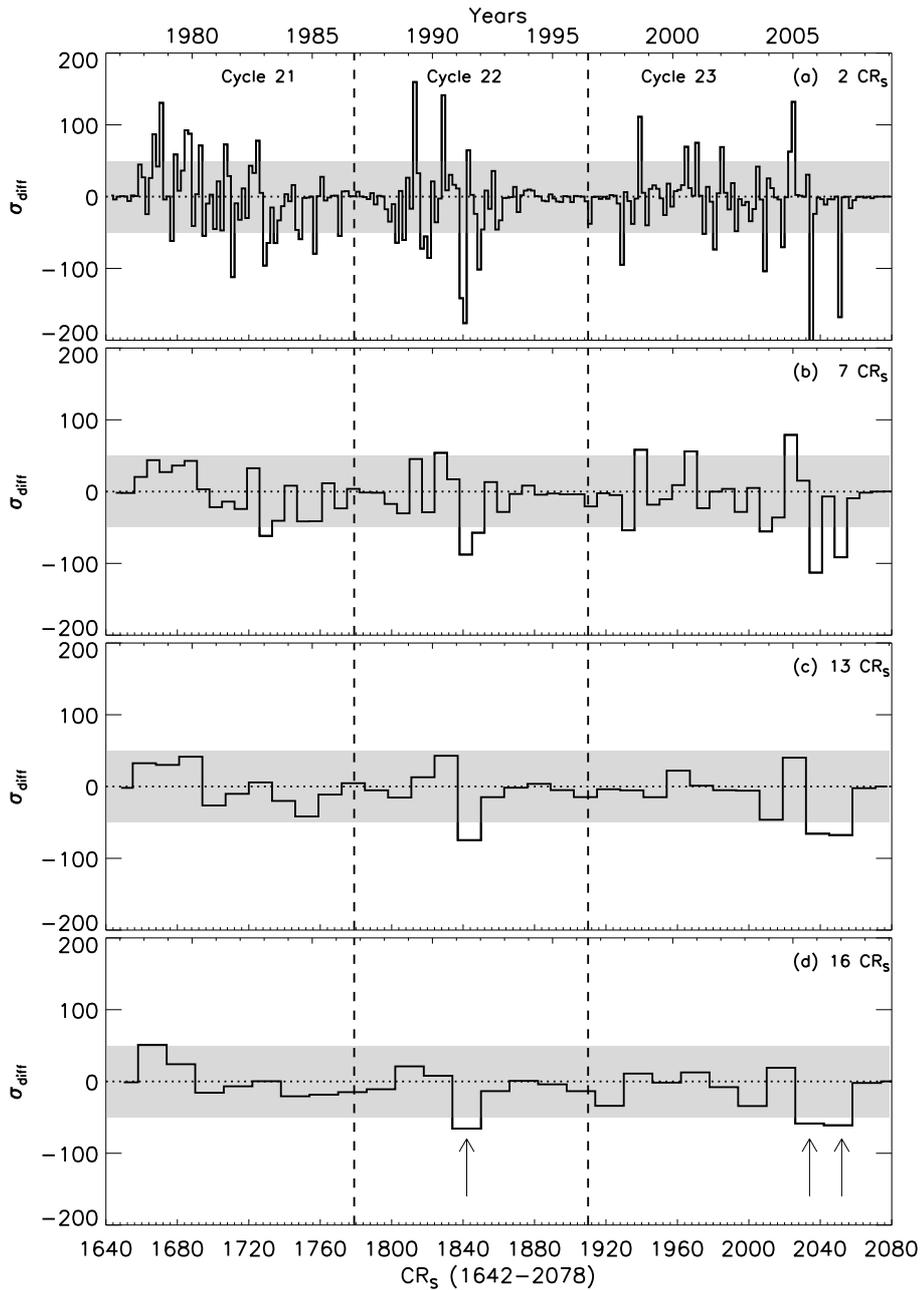}
\caption{A few examples showing the difference of standard deviation of $FI_{SXR}$ ($\sigma_{diff}$) in the northern and the southern hemispheres for four cases with increasing bin sizes of 2, 7, 13, and 16 CRs. Note that $\sigma_{diff}$ exhibited large variations for shorter binning intervals while variations, in general, reduced with increase in the bin size. Gray shaded area denotes $\sigma_{diff}$ below $\pm$50.}  
\label{Fig6}
\end{figure*}

\begin{figure*}
\sidecaption
\includegraphics[width=12cm]{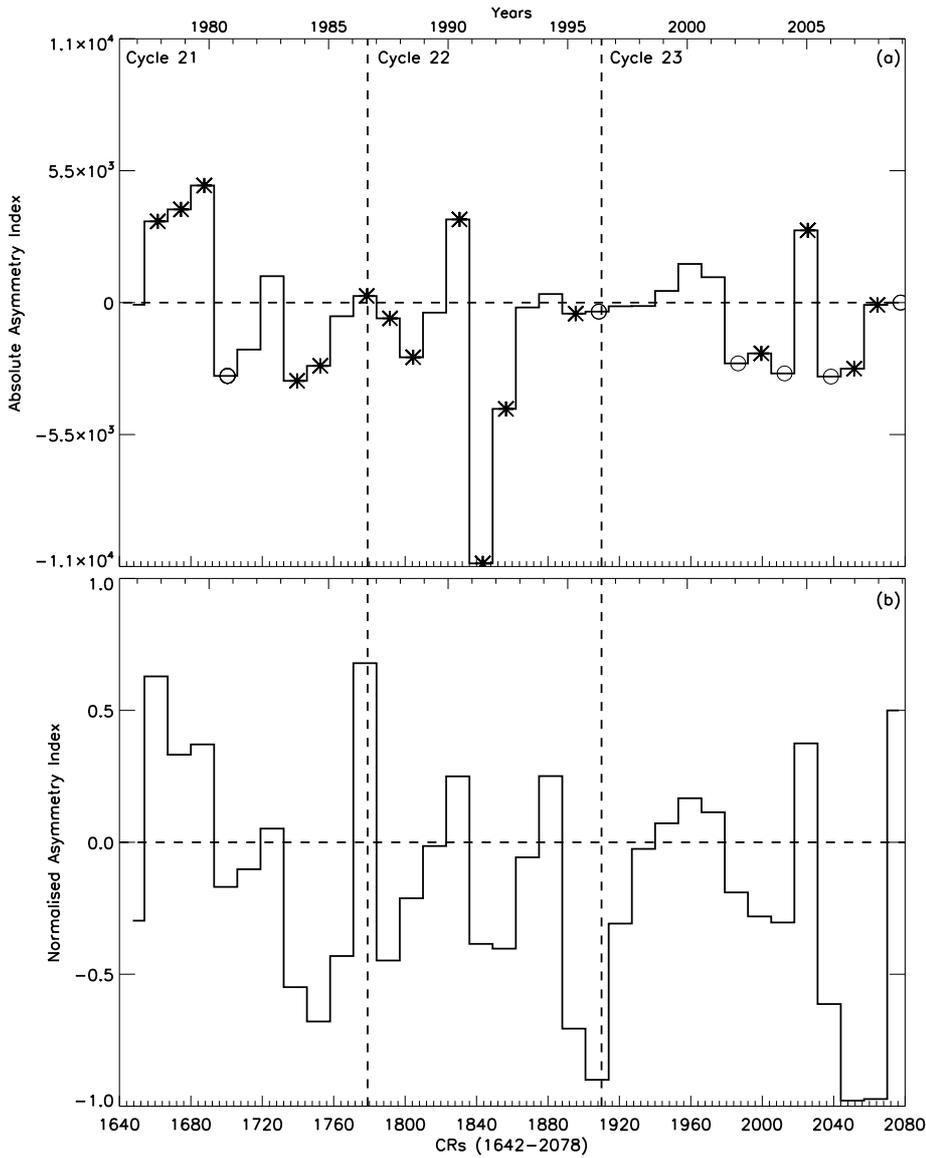}
\caption{North-south (N-S) asymmetry of $FI_{SXR}$ during solar cycles 21, 22, and 23. The asymmetry has been represented by absolute asymmetry index (panel~a) as well as normalized asymmetry index (panel~b). The $FI_{SXR}$ is averaged over 13 Carrington rotations for the asymmetry analysis. Vertical dashed lines indicate solar activity minima. The statistical significance of N-S asymmetry is assessed by Student's t-test and the results are marked by circle/asterisk symbols in top panel. Here asterisks and circles represent N-S asymmetry with significance levels of  $\geq$95 \% and 80--95 \%, respectively.}  
\label{Fig7}
\end{figure*}

\begin{figure*}
\sidecaption
\includegraphics[width=12cm]{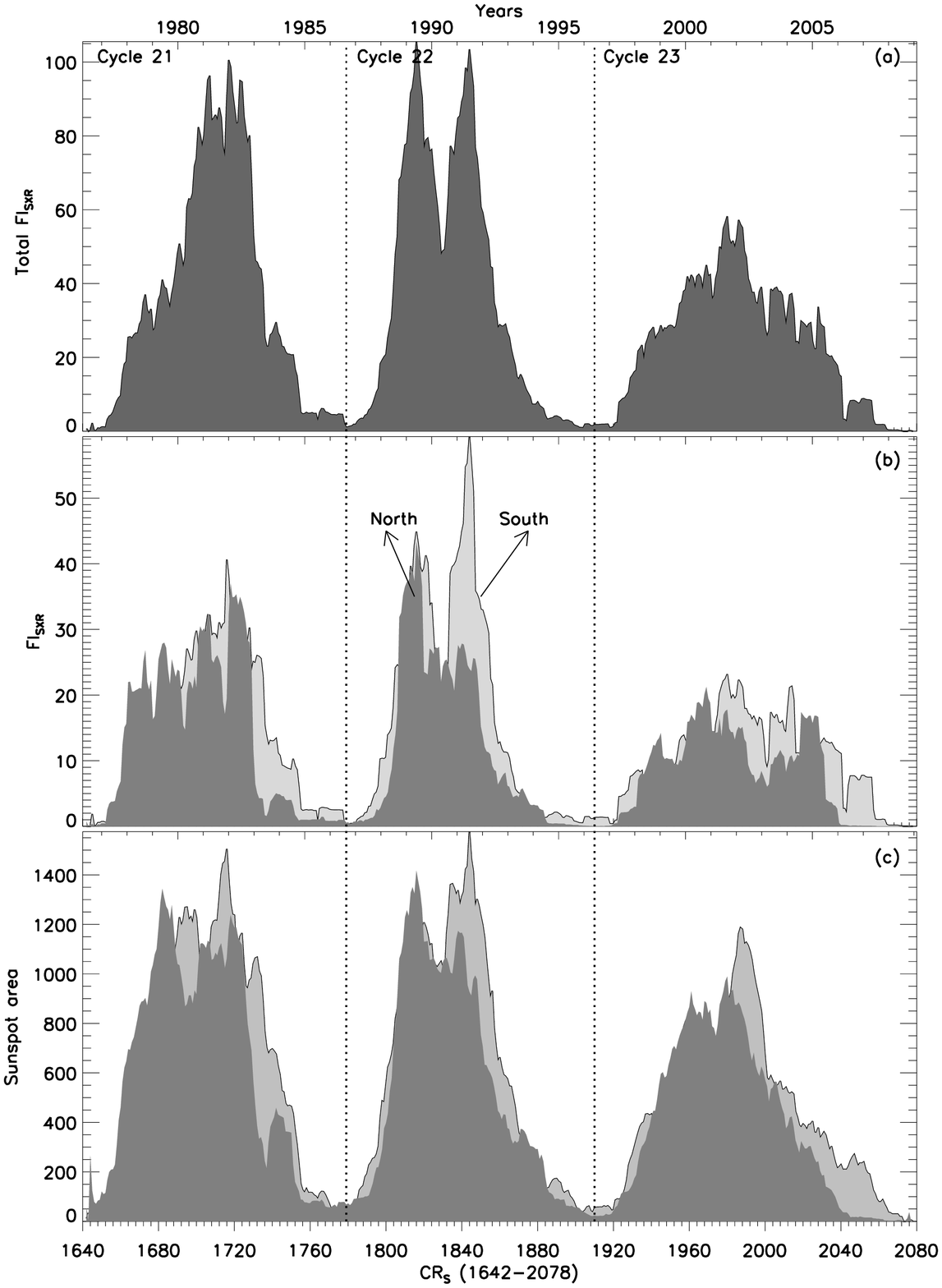}
\caption{(a): Temporal evolution $FI_{SXR}$ is shown by plotting the running average over 13-CR. (b)-(c): The evolution of $FI_{SXR}$ and sunspot area for the northern (black shaded area) and the southern (gray shaded area) hemispheres. The excess of solar activity in one hemisphere over the other can be well understood by these smoothed plots.}
\label{Fig8}
\end{figure*}

In Fig.~\ref{Fig4}, we show rotational average of $FI_{SXR}$ separately for the northern and the southern hemispheres of the Sun. The plots clearly indicate that the flare activity is asymmetric in the two hemispheres. Also noteworthy are the epochs in every cycle when this asymmetry is enhanced. 
The N-S asymmetry is further explored in Fig.~\ref{Fig5} by plotting the cumulative values of $FI_{SXR}$ for solar cycles 21, 22, and 23. For a comparison with sunspot activity, the 
plot is further overlaid with cumulative values of sunspot area. This representation of N-S asymmetry makes it possible to examine the subtle evolutionary aspects of solar cycle, which are missed when 
the activities in the two hemispheres are compared by averaging the activity index over certain time periods \citep{garcia1990}. In these plots, the vertical spacing between the two 
curves at every instant is a measure of the northern/southern excess of the respective activity parameter up to that time.   

To assess the significance of the observed N-S asymmetry, we applied the paired Student's t-test. The test statistics \^{t} is defined by 

\begin{equation}
$\^{t}$=\frac{(\sum D_{i})/n}{\sqrt{\frac{\sum D^{2}_{i}-(\sum D_{i})^{2}/n}{n(n-1)}}},
\end{equation}
where $D_{i}$ is the difference of paired values (here $FI _{SXR}$ in the northern and the southern hemispheres) and $n$ represents the number of elements in 
each group over which the corresponding time series is split. The parameter $n-1$ is the number of degrees of freedom. Let  $({FI_{SXR}})_{N}$ and $({FI_{SXR}})_{S}$ 
denote the averages of $FI_{SXR}$ over a certain period for the northern and the southern hemispheres, respectively. The 
Student's t-statistic and the corresponding probability then essentially signify the two sample populations: $({FI_{SXR}})_{N}$ and $({FI_{SXR}})_{S}$, to have significantly different mean values. The fundamental assumption in applying this test is the data to be selected from populations with the same true variance. 
The Student's t-test is performed  using IDL routine $\it {TM\_TEST}$.

The Student's t-test has been used in previous studies to examine the N-S asymmetry of sunspot distribution \citep[e.g.,][]{temmer2002}. In the present case, we are using this test to evaluate the asymmetry in the distribution of $FI_{SXR}$ in the two hemispheres, which represents the weighted flare intensity over a time interval, while the individual flare events are localized both in space and time. In a strict mathematical sense, the Student's t-test may not be applicable in our case since the data set comprising of flare events in the northern and the southern hemispheres do not have same true variance (i.e., not Gaussian in distribution). Therefore, it is useful to first compare the variance of the $FI_{SXR}$ of the two data sets over different time intervals with increasing bin sizes and then select an optimal time interval for studying the N-S asymmetry of $FI_{SXR}$. For this purpose, we examined the difference of standard deviations of $FI_{SXR}$ in the northern and the southern hemispheres ($\sigma_{diff}=\sigma(FI_{SXR})_{N}-\sigma(FI_{SXR})_{S}$) with increasing bin sizes from 1 CR to 16 CRs in steps of 1 CR. 
Further examination of $\sigma_{diff}$ beyond a bin size of 16 CRs is not found to be statistically  meaningful because of a significant reduction
in number of available data points.
We provide relevant examples of this comparison in Fig.~\ref{Fig6}, where $\sigma_{diff}$ is plotted with bin sizes of 2, 7, 13, and 16  CRs. The plot confirms the  $\sigma_{diff}$ to exhibited large variations for shorter binning intervals while variations, in general, are reduced with an increase in the bin size. With a bin size of 13 CR, the $\sigma_{diff}$ is minimal with a value $<$ 50 except at the three phases (marked by arrows in Fig.~\ref{Fig6}, panel d) which are not unique to this particular case but are
 present with other successive binning intervals also (i.e., 14, 15, and 16). Additionally, binning intervals above 13 CRs does not yield better results, and only decrease the number of samples for the asymmetry analysis within a cycle. We then select 13 CRs to be an optimal timescale to examine the asymmetry explored in this work.  

The N-S asymmetry of $FI_{SXR}$ with 13 CR averaged values is shown in Fig.~\ref{Fig7} using the following two representations: absolute asymmetry index and normalized asymmetry index. The absolute asymmetry index (Fig.~\ref{Fig7}, panel a) gives the difference of values in the northern and the southern hemispheres ($\Delta$$NS$ = ($FI_{SXR})_{N}$$-$($FI_{SXR})_{S}$). This plot also indicates the statistical significance of asymmetry values (i.e., the probability determined by Student's t-test for each data set of 13 CR averaged $(FI_{SXR})_{N}$ and $(FI_{SXR})_{S}$) and the results with probability $\geq$80 \% are annotated by different symbols (asterisks and circles for significance levels of $\geq$95 \% and 80--95 \%, respectively).
 
It is traditional to depict the N-S asymmetry in term of normalized asymmetry index, defined as

\begin{equation}
A_{NS}=\frac{({FI_{SXR}})_{N}-({FI_{SXR}})_{S}}{({FI_{SXR}})_{N}+({FI_{SXR}})_{S}}.
\end{equation}

\noindent In panel b of Fig.~\ref{Fig7}, we have shown the evolution of A$_{NS}$. We compare the evolution of $FI_{SXR}$ for the full solar disk, northern hemisphere, and southern hemisphere by plotting their running averages taken over 13 CRs in Fig.~\ref{Fig8}(a)-(b). In order to study the association of the evolutionary aspects of flare activity with sunspot activity, we also show the running average of sunspot area over 13-CR for the individual hemispheres in this figure (panel~c).

\section{Discussions}
\label{sec_dissus}
We present statistical analysis of temporal evolution and hemispherical asymmetry of soft X-ray flares during solar cycles 21, 22, and 23. Here the flare activity is characterized by $FI_{SXR}$ instead of traditional flare counts. We believe that representation of flare activity in terms of $FI_{SXR}$ is more useful over flare counts as the flare index encompasses information  about flare occurrences over a certain period as well as the weighted intensity of events. 
The flare statistics given in Table~\ref{table:1} highlights the advantage of $FI_{SXR}$. We recognize different perspectives when flare statistics is sampled in terms of their counts and $FI_{SXR}$. For example, X-class flares form a very small fraction of the total number of flares for a given cycle ($\sim$1--1.5~\%), while their contribution in terms of $FI_{SXR}$ is quite significant ($\sim$30--35~\%). On the other hand, C-class flares occur most frequently ($\sim$60--75~\%) but in terms of $FI_{SXR}$ their contribution toward total activity during a cycle is always less than X- or M- class events. 
Notably, the peak intensity of soft X-ray emission can be taken as the proxy for flare's thermal energy content, which is intimately connected to the high energy (non-thermal) processes occurring in the solar corona during a flare. Hence, we believe that it is important to examine the cyclic evolution of flare activity in terms of $FI_{SXR}$ and compare the results with more conventional activity parameters.
Moreover, $FI_{SXR}$ has previously been used with objectives  
to understand the association of flare activities with different solar phenomena of varied spatial and temporal scales \citep[][]{landi1998,park2010,jing2006}.
 
The cross-correlation analysis of $FI_{SXR}$ with other fundamental and well-accepted solar activity parameters reveals that $FI_{SXR}$ has a good positive correlation with them (Fig.~\ref{Fig2} and Table~\ref{table:3}). The comparison of correlation coefficients given in Table~\ref{table:3} clearly indicates that $FI_{SXR}$ has maximum correlation with $FI_{H\alpha}$ ( $r$ = 0.82, cf. Fig.~\ref{Fig2}), which is expected because both indices represent flare activity. However, we should note that during a flare, soft X-ray and H$\alpha$ emissions are produced from distinct parts of a coronal loop threading different layers of the solar atmosphere \citep[for a review, see][]{joshi2012}. The H$\alpha$ emission originates from the chromospheric layers of the Sun where the footpoints of the coronal loops are anchored. On the other hand, SXR emission is produced in coronal loops and indicates the presence of hot plasma ($\sim$10$-$20~MK) inside them. Because of this reason, the size and intensity levels of a flare in H$\alpha$ and SXR may not have one-to-one correspondence. This is also reflected in the scatter plots between $FI_{SXR}$ and $FI_{H\alpha}$, as the correlation coefficients between the two indices vary in different cycles and never exceeds  90\% ($6^{th}$ column of Table~\ref{table:3}).

The correlation of $FI_{SXR}$ with sunspot area is significantly higher than sunspot number (Fig.~\ref{Fig2}). This fact is even more noticeable when the scatter plots of individual cycles are considered (see $3^{rd}$ and $4^{th}$ columns of Table~\ref{table:3}). Here we emphasize that although sunspot number and sunspot area represent same manifestation of solar activity, the spatial location of the underlying process of flux emergence may lead to differences in their activity levels. In a simple sense, sunspot number is a measure of total counts of distinct centres on the solar disk where intensive magnetic flux emergence takes place. Here it is important to note that any variations in sunspot number would depend on the location of the new flux emergence. If the new flux is emerged from locations away from developed sunspot groups the sunspot number would tend to increase. On the other hand, emergence of new flux in the vicinity of, or within, fully developed sunspots may not show corresponding increase in their total numbers, while significantly enhancing active region's magnetic complexity \citep[see, e.g.,][]{ballester2002}. It is well known that the occurrence rate of flares is highly dependent on the magnetic complexity of active region. The significantly high correlation of $FI_{SXR}$ with sunspot area than sunspot numbers is therefore suggestive of the fact that sunspot area is more sensitive toward describing the magnetic complexity, which leads to higher flare productivity.  
We further note that during cycle 23, correlation coefficients of sunspot area and sunspot number with $FI_{SXR}$ show appreciable variations (0.74 vs 0.58).
Here it is important to emphasize that during cycle 23, $FI_{SXR}$ includes a relatively larger fraction of smaller flares (B and C class events) compared to cycles 21 and 22 (cf. Table~\ref{table:1}). 
On the basis of above observations, we speculate that during cycle 23  magnetic flux emergence took place in a more scattered way, resulting the observed deviations in the evolutionary patterns of sunspot number, sunspot area, and flare activity. \cite{ballester2002, ballester2004} studied the periodic variability of photospheric magnetic flux during cycles 21, 22, and 23. Their study provides evidence for a possible link between the spatial evolution of magnetic flux emergence and rate of solar flares, which is consistent with our results \citep[see also][]{joshi2006aa}. In particular, many irregularities in sunspot manifestations during cycle 23 have been well observed and 
extensively studied \citep{russell2010,lukianova2011,lefevre2011,clette2012,livingston2012,chapman2014} that finally point to the complexities of underlying solar dynamo processes. \cite{lefevre2011} found that cycle 23 is quite distinguishable from the previous cycle in terms of significant deficit in small sunspot, while large-scale spots remained largely unaffected. This behavior was pronounced in sunspots with short lifetimes indicating scale-dependent changes in sunspot properties \citep[see also][]{clette2012}. Infrared spectral observations of sunspots from 1998 to 2011 (cycle 23 and beginning of cycle 24) provide crucial evidence for a continuous weakening in the magnetic field strength of sunspots \citep{livingston2012}.

The soft X-ray flare index exhibits a good correlation with solar radio flux observed at 10.7 cm (in fact better than that with sunspot number; cf. Fig.~\ref{Fig2} and Table~\ref{table:3}). This indicates that the short-term variations in $F10.7$, which originates in high chromospheric and low coronal levels, is closely associated with the flare activity. Here we also mention that soft X-ray flare, representing emissions from hot coronal loops, is essentially an indicator of coronal activity.

In Fig.~\ref{Fig3}, we have examined correlations of $FI_{H\alpha}$ with sunspot number and sunspot area. We find that in comparison to $FI_{SXR}$, $FI_{H\alpha}$ exhibits stronger positive correlation with sunspot related parameters (cf. Fig.~\ref{Fig2} (panels a and b) and Fig.~\ref{Fig3}). 
We have already mentioned in the earlier section that the two flare indices represent consequence of the basic energy release process by magnetic reconnection at different altitudes of solar atmosphere. In addition,  although $FI_{SXR}$ indicates good positive correlation with both sunspot parameters, its correlation coefficients between sunspot number and area show significant difference ($r$=0.66 vs 0.78). On the other hand, in case of $FI_{H\alpha}$, the correlation coefficients are comparable ($r$=0.86 vs 0.88). Further, both flare indices exhibit better correlation with sunspot area. This suggests that among the two parameters describing sunspot formation, sunspot area should be considered more `robust' in characterizing solar activity.

A comparison of time profiles of different solar activity parameters, presented in Fig.~\ref{Fig1}, reveals that $FI_{SXR}$ and sunspot area show more fluctuations from their mean level compared to sunspot number and $F10.7$. These plots further show that the level of activity during cycle 23 is lower than previous solar cycles for all the indices. This clearly indicates that Gnevyshev-Ohl (G-O) rule \citep{gnevyshev48} was violated for the pair of cycles 22/23 for all the indicators of solar activity, which are essentially related to different layers of solar atmosphere and computed through multiwavelength measurements. The G-O rule states that the sum of sunspot numbers over an odd cycle exceeds that of the preceding even cycle. This rule is also referred to as the even-odd effect. With the exception of the pair of cycles 4/5, this relationship held until cycle 21. If cycle amplitudes are compared then the pair of cycles 8/9 is also an exception \citep[see, e.g.,][]{hathaway2010}. 

The temporal evolution of sunspot activity (number as well as area) within a cycle reveals the existence of two peaks during the maximum phase of activity, which can easily be recognized from the smoothed plots of Fig.~\ref{Fig1}. The occurrence of double peaks was first recognized by \cite{gnevyshev1967} and the interval between the two peaks is often referred to as the Gnevyshev Gap \citep[see also][]{norton2010}. However, it is noted from Fig.~\ref{Fig1} that the distinctness of the double-peak structure varies between different cycles and activity parameters. We also find that among five parameters considered here, the Gnevyshev peaks are the most prominent in sunspot area with gaps of $\sim$22, $\sim$19, and $\sim$23 months for cycle 21, 22, and 23, respectively. 

The plot of $FI_{SXR}$ in the northern and the southern hemispheres of the Sun clearly indicates the existence of hemispherical asymmetry in the occurrence of solar flares (Fig.~\ref{Fig4}). A further investigation of asymmetrical evolution of $FI_{SXR}$ and its comparison with sunspot area is shown in Fig.~\ref{Fig5} by 
presenting the variation of their cumulative counts. We find that during solar cycle 21 $FI_{SXR}$  exhibits a northern excess, which prevails up to the end of the cycle. In the sunspot area of cycle 21, there is a northern excess during most of the cycle and the temporal variations are also comparable to that of FI$_{SXR}$, however, in the last phase of the cycle sunspot area in the southern hemisphere increased rapidly making a southern excess by the end of the cycle. From this observation, we infer that an increase in sunspot area in the southern hemisphere does not cause enhancement in flare activity on similar scales. During cycles 22 and 23, $FI_{SXR}$ and sunspot area show similar variations. Both cycles exhibit a southern excess by the end of the cycle. Complimentary to the above observations, it is further useful to compare N-S asymmetry of sunspot area with sunspot number. In earlier works, N-S asymmetry of number of sunspots and sunspot groups have been studied in detail \citep{li2001, temmer2002, bankoti2011}. A comparison of our results with these studies clearly shows that the evolution of asymmetry in both sunspot related parameters (area and number) are consistent during solar cycles 21--23.

We evaluated the reliability of the observed N-S asymmetry in $FI_{SXR}$ by applying the Student's t-test. Although this test is widely used to examine the significance of asymmetry between two data sets, it is also useful to recognize an optimal timescales over which the asymmetry can be considered meaningful for the phenomenon under investigation. In our case, a comparison of variance in the distribution of $FI_{SXR}$ between the northern and the southern hemispheres reveals that a N-S asymmetry should be considered meaningful in a global sense when averages are taken over longer time intervals with a comparable true variance being accomplished at 13 CRs (Fig.~\ref{Fig6}).  
Our above finding of $\sim$13 CRs as an optimal interval to examine N-S asymmetry then provides a possible minimum timescale over which the distribution of $FI_{SXR}$ in the two hemispheres are randomized enough for the t-test to be applicable. The timescale of $\sim$13 CRs, being approximately 1 calendar year, agrees well with the timescale considered in some of the N-S asymmetry investigations \citep[see, e.g.,][]{atac1996,joshi2005aa,gao2009}.

In Fig.~\ref{Fig7}, we have shown N-S asymmetry in terms of absolute and normalized asymmetry indices. The Student's t-test reveals that the observed asymmetry is highly significant in 65\% cases (see Fig.~\ref{Fig7}, panel a). In the representation of normalized asymmetry (Fig.~\ref{Fig7}, panel b), the large asymmetries at minimum phases is a trivial consequence of occurrence of small number of flares during solar cycles minima. 
In Fig.~\ref{Fig8}, we plot 13-CR running average of $FI_{SXR}$ and sunspot area for the two hemispheres. From this representation, it is evident that the dominance of one hemisphere over the other persists for several rotations. In general during cycles 21--23, the northern hemisphere was more active during the ascending phase of the cycle, while the southern hemisphere exhibits dominant activity during the descending phase. The successive dominance of the northern and the southern hemispheres during ascending and descending phases, respectively, in the last three cycles has earlier been reported in other solar activity indicators such as, sunspot number, sunspot area, H$\alpha$ flare index, photospheric magnetic flux etc. \citep [see, e.g., ][]{atac1996,temmer2006,li2009,vernova2014}. This plot also reveals information about the origin of the dual peaked structure of a solar cycle, which can be recognized for all the solar parameters plotted in Fig.~\ref{Fig1}. We find that the dual peak structures correspond to the amplitude of activity in different hemispheres (Fig.~\ref{Fig8}, panels b and c). 

\section{Conclusions}
\label{sec_con}
We have characterized solar activity in terms of soft X-ray flare index ($FI_{SXR}$) instead of the traditional flare counts and compared the results with conventional solar activity indices. The analysis of $FI_{SXR}$ reveals many subtle aspects of solar activity evolution as, in addition to flare counts, the $FI_{SXR}$ relates to the peak SXR intensity of individual events. In the following, we highlight the important results of this study:

\begin{enumerate}

\item
The study reveals that $FI_{SXR}$ correlates well with other conventional parameters of solar activity. Further, it is noted that among two parameters describing sunspot activity, namely, sunspot number and sunspot area, $FI_{SXR}$ shows significantly higher correlation with sunspot area over sunspot number, which suggests the variations in sunspot area to be more closely linked with the transient energy release in the solar corona.

\item
The cumulative plots of $FI_{SXR}$ for the northern and the southern
hemispheres indicate N-S excess through various phases of solar cycles. It is found that, in terms of $FI_{SXR}$, overall there was a slight excess of activity in the northern hemisphere for cycle 21, while a southern excess prevailed for cycles 22 and 23.

\item
The study reveals a statistically significant N-S asymmetry in $FI_{SXR}$ which becomes more reliable and consistent with the data being binned over longer periods. By comparing the variance of $FI_{SXR}$ in the two hemispheres, we obtain a period of 13 CR as an optimal time interval
to explore N-S asymmetry in intensity weighted flare index. In view of this, a period of $\sim$13 CRs can be regarded as a possible minimum timescale over which the distribution of $FI_{SXR}$ in the two hemispheres are randomized enough to manifest a significant N-S asymmetry.

\item
During cycles 21--23, generally the northern hemisphere was more active during the ascending phase of the cycle, while the southern hemisphere dominated during the descending phase. 
\end{enumerate}

These results imply that the evolution of flare activity in the northern and the southern hemispheres of the Sun exhibits significant asymmetry. Importantly, the N-S asymmetry of solar flares is manifested with respect to the occurrence of flares along with the intensity of flare events. The observations of various phenomena of N-S asymmetry is crucial as these may provide valuable constraints for solar dynamo models.

\begin{acknowledgements}
We acknowledge National Geophysical Data Centre (NGDC), Royal Greenwich Observatory, and Kandilli Observatory for providing the required data used in this study. This work was supported by the BK21 plus program through the
National Research Foundation (NRF) funded by the Ministry of Education of 
Korea. We sincerely thank an anonymous referee for
providing constructive comments and suggestions toward enhancing the quality and presentation of this paper.
\end{acknowledgements}


\end{document}